\documentclass[10pt]{amsart}

\usepackage{amssymb}
\usepackage{amsmath}
\usepackage{amsthm} 
\usepackage{mathrsfs}                                                           
\usepackage{slashbox}  
\usepackage[all]{xy}
\usepackage{wrapfig} 
\def\2{{1\over 2}}
\def\Tr{{\rm Tr}}

\newcommand{\rf}[1]{(\ref{#1})}
\def\b{\bar}

\newcommand{\ud}{\mathrm{d}}
\renewcommand{\t}{\tilde}
\newcommand{\p}{\partial}

\newcommand{\mA}{\mathbf{A} }
\newcommand{\mB}{\mathbf{B} }
\newcommand{\mC}{\mathbf{C} }

\newcommand{\mV}{\mathbf{V} }
\newcommand{\mW}{\mathbf{W} }

\newcommand{\mF}{\mathcal{F} }
\newcommand{\mQ}{\mathcal{Q} }
\newcommand{\m}{\ud*\ud}

\newcommand{\mg}{\mathfrak{g}}

\newcommand{\mG}{\mathbf{G}}
\newcommand{\mK}{\mathfrak{K}}
\newcommand{\md}{ \mathfrak{D}}
\newcommand{\tmd}{{\mathbb{D}}}
\newcommand{\mFs}{\mathcal{F}_{susy} }
\newcommand{\mKs}{\mathcal{K}_{susy} }
\title{\bf{Homotopy Algebras of Differential (Super)forms in Three and Four Dimensions}}
\author{Martin Rocek}
\author{Anton M. Zeitlin}
\address{\newline 
Martin Rocek,\newline
C.N. Yang Institute for Theoretical Physics,\newline
Stony Brook University,\newline
Stony Brook, NY 11794-3840, USA\newline
rocek@insti.physics.sunysb.edu, \newline
http://insti.physics.sunysb.edu/$\sim$rocek\newline}
\address{\newline 
Anton M. Zeitlin,\newline
Department of Mathematics,\newline
Louisiana State University,\newline
303 Lockett Hall,\newline
Baton Rouge, LA 70803-4918, USA;\newline
IPME RAS, V.O. Bolshoj pr., 61, 199178,\newline 
St. Petersburg\newline
zeitlin@lsu.edu,\newline
http://math.lsu.edu/$\sim$zeitlin \newline  
http://www.ipme.ru/zam.html }

\begin{document}
\maketitle
\begin{abstract}
We consider various $A_{\infty}$-algebras of differential (super)forms, which are related to gauge theories and demonstrate explicitly how certain reformulations of gauge theories lead to the transfer of the corresponding $A_{\infty}$-structures. In addition, for $N=2$ 3D space we construct the homotopic counterpart of the de Rham complex, which is related to 
the superfield formulation of the $N=2$ Chern-Simons theory. 
\end{abstract}

\\

\section{Introduction}
Homotopy associative algebra, or $A_{\infty}$-algebra is a certain generalization of a differential graded associative algebra. For the first time this concept appeared in the 1960s in the works of Jim Stasheff \cite{stasheff}, \cite{stashbook} in the context of Algebraic Topology. The theory continued developing both in the 1970s and the 1980s, but it wasn't until the 1990s that people realized that the same structures appeared in String Physics and Geometry, e.g. in the context of String Field Theory \cite{zwiebach} and Mirror Symmetry \cite{konts}. In this article, we are studying several examples of such algebraic objects related to the basic gauge (Super) Field Theories, following previous results obtained in  \cite{lzgauge}, \cite{azjpa}, \cite{azcomm},  and related articles devoted to $L_{\infty}$-algebras, e.g. \cite{sftz}, \cite{azjhep}, \cite{azijmpa}, see also  \cite{movshev} and recent paper \cite{hohm}. We also note that there is an alternate approach to the homotopy 
algebras, associated to the dynamical systems, known as unfolded formalism, which first appeared in the context of higher spin theories, see e.g. \cite{vas1}, \cite{vas2}. 

First, let us give an idea of what $A_{\infty}$-algebra is and present one of the reasons why it is a natural object from the mathematical point of view. Suppose, $\mF$ is a differential graded algebra (DGA), i.e. $\mF$ is a chain complex that has an associative bilinear operation satisfying the Leibniz rule with a differential. Assume, there is another complex $\mK$, homotopically equivalent to $\mF$. Therefore, we can deduce that the cohomology of $\mK$ is isomorophic to $\mF$, and both have the structure of an associative algebra. The natural question is the following: what happens on the level of chain complexes? It appears that we have a transfer of the algebraic structure from $\mF$ to $\mK$ but in general the transferred product will not be associative. The structure induced from the bilinear operation and the homotopical equivalence is what we call $A_{\infty}$-algebra \cite{gug}, \cite{merk},\cite{konts}. The induced bilinear operation $\mu_2$ on $\mF$ will satisfy an associativity condition up to homotopy that is described by a trilinear operation $\mu_3$. Moreover, 
(see e.g.~\cite{merk},\cite{konts},\cite{markltransf},\cite{hu},\cite{keller}) one can continue and construct operations $\mu_n$, satisfying ``higher" bilinear associativity relations. Thus, if we are given the operations $\mu_n$, $n=(1, 2,3, ...)$ satisfying such bilinear relations, we say that there is the structure of $A_{\infty}$-algebra on $\mK$. In fact, all these relations can be summarized in one: $\p^2=0$ (see Appendix A for details). 

An important feature of the $A_{\infty}$-algebras is that one can define a generalization of the generalized Maurer-Cartan (GMC) equation: 
\begin{eqnarray}
Q\Phi+\mu_2(\Phi,\Phi)+\mu_3(\Phi, \Phi,\Phi)+...=0,
\end{eqnarray}
where $\Phi$ has degree $1$. This equation possesses the following infinitesimal symmetry transformation:
\begin{eqnarray}
&&\Phi\to Q\Lambda+\mu_2(\Phi,\Lambda)-\mu_2(\Lambda,\Phi)+\nonumber\\
&&\mu_3(\Lambda,\Phi,\Phi)-\mu_3(\Phi,\Lambda,\Phi)+\mu_3(\Phi,\Phi,\Lambda)...,
\end{eqnarray}
where $\Lambda$ is infinitesimal and has degree 0.  

In physics, particularly in String Field Theory, $A_{\infty}$-algebras appear in the context of the Batalin-Vilkovisky (BV) formalism (see e.g.~\cite{kajiura}). Namely, for a given action functional satisfying the BV master equation, one can construct the nilpotent noncommutative odd vector field $\mathbf{Q}$ encoding all the operations of the $A_{\infty}$-algebra (see Appendix A).

In this article, we will be studying homotopy algebras related to (super)field theories, as well as their transfers. 

In Section 2, as an invitation to the subject, we examine the $N=2$  D=3 Chern-Simons theory in the superfield formulation (this theory appears to be physically relevant, see e.g. \cite{jafferis},\cite{klebanov}). Using the corresponding BV action, we 
construct the superfield counterpart of the de Rham Complex in three dimensions that turns out to be a homotopy algebra instead of the usual DGA.

In Section 3, we consider the Yang-Mills (YM) theory in its usual second and  first order formulations. We describe the corresponding $A_{\infty}$-algebras, for which GMC coincides with YM equations 
of motion, and show how they are related to the Batalin-Vilkovisky (BV) formalism. 
In this case, the first order formulation gives us just a differential graded algebra, while the second order formulation leads to an $A_{\infty}$-algebra where all the operations starting from the quadrilinear one vanish.

In the end of Section 3, we are showing that these two $A_{\infty}$-algebras are related by the transfer formula. This is one of the simplest non-artificial examples of the transfer formula. 
This way, we observe that some of the basic field theory reformulations, like the transfer from the second order action to the first order action, 
can be explained purely on the level of homotopical algebra (see also \cite{sftz}). 

In Section 4, we generalize these results to the supersymmetric case. We use the superfield formulation of 
$N=1$ supersymmetric (SUSY) YM theory and study the homotopical algebras corresponding to the second and first order formulations. In this case, $A_{\infty}$-algebras are ``richer", i.e. we have an infinite chain of polylinear operations in both first and second order formulations. As in the previous example, they are related by the transfer formula.
In the end, we discuss some further developments and future tasks, in particular the meaning of the representation of real superfield via complex ones from the homotopical algebra point of view. \\

\noindent {\bf Acknowledgements.} We are grateful to K. Costello, M. Markl, M. Movshev and J. Stasheff for illuminating discussions. We are indebted to the valuable comments of the referee. 
We would like to express our gratitude to the wonderful environment of  Simons Summer Workshops where this work was partially done. A.M.Z. is grateful to A.N. Fedorova for careful reading of the manuscript. M.R. is happy to thank NSF grant PHY-1620628 for support.
A.M.Z. is supported by AMS Simons travel grant.

\section{ Homotopic de Rham complex in $N=2$ 3D superspace.}
\noindent{\bf 2.1. De Rham complex in 3D and BV Chern-Simons.} Consider the DGA related to the de Rham complex in three dimensions:
\begin{eqnarray}
0\to\Omega^0\xrightarrow{d}\Omega^1\xrightarrow{d}\Omega^2\xrightarrow{d}\Omega^3\to 0,
\end{eqnarray}
where $\Omega^i$ is the space of differential forms of degree $i$. Moreover, this complex possesses a natural 
pairing given by the integral of the wedge product of forms and trace, giving a cyclic structure of the de Rham DGA. The same DGA structure can be constructed for the forms with values in $U({\mg})$ where $\mg$ is some semisimple Lie algebra. 
Then the formal action functional, associated with this cyclic DGA (see Appendix A), corresponding to the element $\Phi=c+\mA+\mA^*+c^*$, where 
$c\in \Omega^{0}[1]\otimes{\mg}$, $\mA\in \Omega^1\otimes{\mg}$, $\mA^*\in \Omega^2[-1]\otimes{\mg}$, $c^*\in\Omega^3[-2]\otimes{\mg}$ has the following form:
\begin{eqnarray}
S[\Phi]=S_{CS}[\mA]+\int d^3x\Tr(d_{\mA}c\wedge \mA^*+[c,c]c^*),
\end{eqnarray}
where index in square brackets stands for the shift of grading to make $\Phi$ an element of degree 1, $d_{\mA}=d+[\mA,\cdot]$ is a covariant derivative, ${\rm Tr}$  stands for the Killing form on  
$\mathfrak{g}$ and  
$S_{CS}[\mA]=\int d^3x ~{\rm Tr}(\mA \wedge d\mA+\frac{1}{3}[\mA,\mA]\wedge\mA)$ is the usual 3D Chern-Simons action. \\

\noindent {\bf 2.2. SUSY de Rham complex and SUSY Chern-Simons.} Let us consider N=2 3D Euclidean superspace. 
This is a superspace where in addition to three even coordinates there are 4 odd coordinates that are the components of Weyl spinors $\theta^{\alpha}$, $\bar\theta^{{\alpha}}$. As in the N=1 4D case, one can consruct  the superderivatives $D_{\alpha}$, $\b D_{\alpha}$. The difference is that one can make new Lorentz scalars out of them, like $D_{\alpha}\b D^{\alpha}$ (see Appendix B.3 for details). 
The relations between superderivatives allow to construct the following complex:
\begin{eqnarray}\label{susydr}
0\xrightarrow{ }\Theta\xrightarrow{id}\Sigma
\xrightarrow{\b D^{\alpha}D_{\alpha}}\t \Sigma\xrightarrow{{\b D}^2}\t \Theta\to 0.
\end{eqnarray} 
Here $\Theta\cong\t \Theta$ is the space of chiral scalar fields (if $\Lambda \in \Theta$, then $\b D_{\alpha}\Lambda=0$), and $\Sigma\cong\t\Sigma$ is the space of complex scalar fields. 
Afterwards, we will denote this complex $(SdR^{\cdot}, d_{S})$. 
It is not hard to show that the 3D de Rham complex can be embedded in \rf{susydr}. Moreover,  one can define a nondegenerate pairing on \rf{susydr} similar to $\langle\cdot, \cdot \rangle_\mF$:
\begin{eqnarray}
&&\langle\cdot, \cdot \rangle: SdR^{i}\otimes SdR^{3-i}\to \mathbb{C},\nonumber\\
&&\langle{\Lambda,\t\Lambda}\rangle=\int d^3 xd^2\theta \Lambda\t \Lambda,\quad \langle V,\t V\rangle=\int d^3 x d^4\theta V\t V,\nonumber
\end{eqnarray}
where $\Lambda\in SdR^{0}$, $V\in SdR^{2}$, $\t V\in SdR^{2}$, $\t \Lambda\in SdR^{3}$. This pairing satisfies  
the  familiar property \rf{cycpairing}. Therefore, we can hope for the existence of a cyclic $A_{\infty}$-algebra related to the complex \rf{susydr}. 
In order to construct it, we can consider the following action giving the superfield formulation to the N=2 Chern-Simons theory \cite{zupnik}:
\begin{eqnarray}\label{zup}
S_{susyCS}=\int d^3 xd^4\theta\int_0^1dt{\rm Tr}\Big (V\b D^{\alpha}(e^{-tV}D_{\alpha}e^{tV})\Big ),
\end{eqnarray}
where $V\in \Sigma\otimes \mathfrak{g}$. This action has the symmetry that is the same for all the supersymmetric theories we are considering here, namely  
\begin{eqnarray}
e^V\to e^{\Lambda}e^Ve^{\bar \Lambda},
\end{eqnarray}
where $\Lambda, \b \Lambda$  are respectively chiral and atichiral scalar fields with values in Lie algebra $\mathfrak{g}$. The corresponding infinitesimal version of the symmetry is:
\begin{eqnarray}
&&V\to V+\delta_{\Lambda, \b \Lambda}V=V+\frac{1}{2}L_{V}(\Lambda-\bar \Lambda+\coth(\frac{1}{2}L_V)(\Lambda+\b \Lambda)),
\end{eqnarray} 
where $L_V\cdot=[V,\cdot]$. Let us restrict the symmetry to the chiral transformations: $e^V\to e^{\Lambda}e^V$. Then, the BV modification of the action \rf{zup} is (cf. \rf{bvsymact}):
\begin{eqnarray}
S^{BV}_{susyCS}=S_{susyCS}+\int d^3 x d^4 \theta {\rm Tr}(\delta_C (V) V^*)+
\int d^3x d^2\theta{\rm Tr} ([C,C]C^*),
\end{eqnarray}
where $C\in SdR^{0}[+1]$, $V\in SdR^{2}$, $V^*\in SdR^{2}[-1]$, $C^*\in SdR^{3}[-2]$. This action will generate the odd vector field that according to the results of Appendix B generates the $A_{\infty}$-algebra. It is obvious that the corresponding chain complex, where the $A_{\infty}$ operations act, is the one from \rf{susydr}. 
Therefore, we have the following proposition. \\

\noindent {\bf Proposition 2.1.} {\it Complex $(SdR^{\cdot}, d_{S})$ possesses a nontrivial $A_{\infty}$ structure (with all the operations nonvanishing) provided by the action functional $S^{BV}_{susyCS}$}.\\

\noindent Let us express the bilinear operation explicitly:

\begin{center}
$\nu_2(f_1,f_2)$=\\
\vspace{5mm}
\begin{tabular}{|l|c|c|c|r|}
\hline
\backslashbox{$ f_2$}{$f_1$}&          $\Lambda_1$ & $V_1$ & $\t V_1$ & $\t\Lambda_1$ \\
\hline
$\Lambda_2$ &                  $\Lambda_1\Lambda_2$    &$\frac{1}{2}V_1\Lambda_2$ & 
$\frac{1}{2}\t V_1\Lambda_2$ &$\t\Lambda_1\Lambda_2$\\
\hline
$V_2$     &  $\frac{1}{2}\Lambda_1V_2 $ & $(V_1,V_2)_h$   &  $-\frac{1}{2}\b D^2(\t V_1V_2)$  &0\\
                    
\hline
$\t V_2$ & $\frac{1}{2}\Lambda_1\t V_2$ &  $-\frac{1}{2}\b D^2(V_1\t V_2)$ &    0&0\\
\hline
$\t\Lambda_2$                     & $\Lambda_1\t\Lambda_2$ &   0 &    0&0\\
\hline
\end{tabular}\\
\end{center}
\noindent where $\Lambda_1, \Lambda_2\in SdR^0$, $V_1, V_2\in SdR^1$, $\t V_1, \t V_2\in SdR^2$, $\t \Lambda_1, \t \Lambda_2\in SdR^3$, and 
\begin{eqnarray}
(V_1,V_2)_h=-\frac{1}{2}D_{\alpha}V_1\b D^{\alpha}V_2-\frac{1}{2}\b D^{\alpha}V_1D_{\alpha}V_2.
\end{eqnarray}
{\bf Corollary 2.1.} {\it The operation $\nu_2$ is homotopy associative on $(SdR^{\cdot}, d_{S})$}. \\

\noindent The equations of motion for $S_{susyCS}$ are \cite{zupnik}:
\begin{eqnarray}
f(L_V)\b D^{\alpha}(e^{-V}D_{\alpha}e^V)=0,
\end{eqnarray}
where $f(x)=\frac{e^{x}-1}{x}$. 
This equation is the GMC equation for the resulting $A_{\infty}$-algebra, which is the 
superfield generalization of the Maurer-Cartan equation in 3D: $d\mA+\mA\wedge\mA=0$. \\

\noindent{\bf Remark 2.1.} Also, we note that in principle, one can extend the complex \rf{susydr} by means of the space of antichiral scalars (compare to the SUSY Yang-Mills case in subsection 3.4.). The resulting complex will be 
\begin{eqnarray}
\xymatrixcolsep{40pt}
\xymatrixrowsep{3pt}
\xymatrix{
0\ar[r]&\Theta \ar[r]^{id} & \Sigma \ar[r]^{\b D^{\alpha}D_{\alpha}}  &\tilde\Sigma \ar[r]^{\b D^2}\ar[ddr]^{D^2} & \tilde\Phi\ar[r]&0\\
 &\oplus&&&\oplus
 &&          &  &&\\
 &\bar{\Theta}\ar[uur]^{-id}&   &&\tilde{\bar{\Theta}}}\nonumber
\end{eqnarray}

The resulting $A_{\infty}$-algebra, generated by the action
\begin{eqnarray}
&&S^{fullBV}_{susyCS}=S_{susyCS}+\int d^3 x d^4 \theta ~{\rm Tr}(\delta_{C,\b C} (V) V^*)+\nonumber\\
&&\int d^3x d^2\theta ~{\rm Tr}([C,C]C^* +[\b C,\b C]\b C^*).
\end{eqnarray}
 will describe the full symmetry of the action \rf{zup}.

\section{Homotopy algebras of Yang-Mills theory.}
\noindent {\bf 3.1. The $A_{\infty}$-algebra of the Maxwell complex \cite{lzgauge}.}
We consider Maxwell complex
\begin{eqnarray}
0\to\mathcal{F}^{0}\xrightarrow{\mathcal{Q}}\mathcal{F}^{1}\xrightarrow{
\mathcal{Q}}
\mathcal{F}^{2}\xrightarrow{\mathcal{Q}}\mathcal{F}^{3}\to 0,
\end{eqnarray}
where the spaces $\mathcal{F}^i$ and the action of $\mQ$ are as follows:
\begin{eqnarray}\label{maxwell}
0\xrightarrow{ }\Omega^{0}(M)\xrightarrow{\ud}\Omega^{1}(M)
\xrightarrow{\m}\Omega^{3}(M)\xrightarrow{\ud}\Omega^{4}(M)\to 0,
\end{eqnarray}
where $M$ stands for any four dimensional (pseudo)Riemannian manifold (chain complex \rf{maxwell} was studied also in \cite{gover}, \cite{waldron} in a different context). 
This complex has a structure of homotopy commutative associative algebra \cite{lzgauge}. Let us introduce the corresponding multilinear operations:
\begin{eqnarray}
&&(\cdot, \cdot)_h: \mathcal{F}^i\otimes \mathcal{F}^j\to \mathcal{F}^{i+j},\nonumber\\
&&(\cdot, \cdot, \cdot)_h: \mathcal{F}^i\otimes \mathcal{F}^j\otimes \mathcal{F}^k\to 
\mathcal{F}^{i+j+k-1}.
\end{eqnarray}

The bilinear operation is defined by means of the following table:\\

$(f_1,f_2)_h$=
\begin{tabular}{|l|c|c|c|r|}
\hline
\backslashbox{$ f_2$}{$f_1$}&\makebox{$v$} & \makebox{$\mA$} & \makebox{$\mV$} & \makebox{$a$} \\
\hline
$w$ & $vw$& $\mA w$& $\mV w$&$a w$\\
\hline
$\mB$ &$v\mB$  & $(\mA,\mB)$   &   $\mB\wedge\mV$&0\\
\hline
$\mW$ & $v\mW$ &  $ \mA\wedge\mW$ &    0&0\\
\hline
$b$ & $vb$ &   0 &    0&0\\
\hline
\end{tabular}\\

\noindent where $f_1$ takes values in the set $\{v, \mA, \mV, a\}$ from the first row, and $f_2$ takes values in the set $\{w, \mB, \mW, b\}$ from the first column. Other elements in the table represent the value of bilinear operation 
$(f_1,f_2)_h$ for appropriately chosen $f_1$ and $f_2$. In the table above $v,w\in\mathcal{F}^0$; 
$\mA,\mB\in \mathcal{F}^1$; $\mV,\mW\in\mathcal{F}^2$; $a,b\in\mathcal{F}^3$. 
The bilinear operation $(\mA,\mB)$ is defined as follows:
\begin{eqnarray}
&&(\mA,\mB)=\mA\wedge(*\ud\mB)-(*\ud\mA)\wedge\mB+\ud *(\mA\wedge \mB).
\end{eqnarray}
The operation $(\ \cdot, \ \cdot, \ \cdot)_h$ is defined to be nonzero only when all the arguments belong to $\mF^1$. For $\mA$, 
$\mB$, $\mC$$\in \mF^1$ we have: 
\begin{eqnarray}
(\mA,\mB ,\mC )_h=\mA\wedge*(\mB\wedge\mC)-(*(\mA\wedge\mB))\wedge\mC.
\end{eqnarray}
By the direct calculations (see \cite{lzgauge}), one can show that these operations satisfy the following relations providing the structure of homotopy commutative $A_{\infty}$-algebra. \\

\noindent{\bf Proposition 3.1.} {\it Let $a_1,a_2, a_3, a_4,b, c$ $\in$ $\mF$. Then the following relations hold:
\begin{eqnarray}
&&\mQ(a_1,a_2)_h=(\mQ a_1,a_2)_h+(-1)^{n_{a_1}}(a_1,\mQ a_2)_h,\nonumber\\
&&(a_1,a_2)_h=(-1)^{n_{a_1}n_{a_2}}(a_2,a_1)_h,\nonumber\\
&&\mQ(a_1,a_2, a_3)_h+(\mQ a_1,a_2, a_3)_h+(-1)^{n_{a_1}}(a_1,\mQ a_2, a_3)_h+\nonumber\\
&&(-1)^{n_{a_1}+n_{a_2}}( a_1, a_2, \mQ a_3)_h=((a_1,a_2)_h,a_3)_h-(a_1,(a_2,a_3)_h)_h
,\nonumber\\
&&(-1)^{n_{a_1}}(a_1,(a_2,a_3,a_4)_h)_h+((a_1,a_2,a_3)_h,a_4)_h=\nonumber\\
&&((a_1,a_2)_h,a_3,a_4)_h-(a_1,(a_2,a_3)_h, a_4)_h+(a_1,a_2,(a_3,a_4)_h)_h,
\nonumber\\
&&((a_1,a_2, a_3)_h,b,c)_h=0.
\end{eqnarray} }
If we tensor complex $(\mF^{\bf\cdot},\mQ)$ with some universal enveloping algebra for some reductive Lie algebra $\mg$, one obtains that inherited operations $(\ \cdot,\ \cdot)_h$ and $(\ \cdot,\ \cdot, \ \cdot)_h$ satisfy the relations of  $A_{\infty}$-algebra on the resulting complex $(\mF_{\mg}^{\bf\cdot},\mQ)$. 

If the manifold $M$ is compact, or the fields are with the compact support, one can show that the Maxwell complex possesses a pairing 
\begin{eqnarray}
\langle f_1,f_2\rangle=\int_M Tr(f_1\wedge f_2),
\end{eqnarray}
 which makes the $A_{\infty}$-algebra introduced above to be cyclic. 
Namely, one can define
\begin{equation}
\{\cdot,..., \cdot\}_h: \mF_{\mg}\otimes ...\otimes \mF_{\mg}\to \mathbb{C} 
\end{equation}
in the following way:
\begin{eqnarray}
&&\{f_1,f_2\}_h=\langle \mQ f_1, f_2\rangle, \quad \{f_1,f_2,f_3\}_h=\langle (f_1, f_2)_h,f_3\rangle,\nonumber\\
&&\{f_1,f_2,f_3, f_4\}_h=\langle (f_1, f_2,f_3)_h,f_4\rangle.
\end{eqnarray}
For more details about the cyclic structures see Appendix A. One of the most important applications of the cyclic structure is that one can write the following action functional in the form that is the ``homotopy'' generalization of the Chern-Simons action functional 
\begin{eqnarray}
S_{HCS}[f]=\frac{1}{2}\{f,f\}+\frac{1}{3}\{f,f,f\}+\frac{1}{4}\{f,f,f,f\}
\end{eqnarray}
such that $f\in\mF_{\mg}^1$ and the variation of this functional with respect to $f$ leads to the generalized Maurer-Cartan equation for $f$.
The Maurer-Cartan equation and its symmetries in the case of the $A_{\infty}$-algebra considered above
\begin{eqnarray}
\mQ\mA+(\mA,\mA)_h+(\mA,\mA,\mA)_h=0\nonumber\\
\mA\to \mA+\epsilon (\mQ u +(u,\mA)_h-(\mA,u)_h),
\end{eqnarray}
which lead to Yang-Mills equations and the gauge symmetry if $f=\mA$, where $\mA\in\mF_{\mg}^1$ and $u\in\mF^0_\mg$.

In the case when $f=c+\mA+\mA^*+c^*$, where $c\in\mF_{\mg}^0[1]$, $\mA\in \mF_{\mg}^1$, $\mA^*\in \mF_{\mg}^2[-1]$,  $c^*\in\mF^3[-2]$, such that the grading of the element from $\mF^i_{\mg}[j]$ is $i+j$,
the action above leads to the BV Yang-Mills action:
\begin{eqnarray}
S_{BVYM}=\int_M\Tr({\bf F}\wedge *{\bf F}+\mA^*\wedge \ud_{\mA}c+[c,c]c^*).
\end{eqnarray}
It is well-known that this action satisfies the so-called BV master equation that, according to the general principle, leads to the $A_{\infty}$-algebra on the complex $(\mF^{\cdot},\mQ)$ (see Appendix A). 
%This $L_{\infty}$-algebra can be obtained form the $A_{\infty}$-algebra which we introduced in this section. Considering appropriate antisymmetrization (see \cite{ladamarkl} and Appendix),
 %one obtains a YM homotopy Lie algebra, e.g. it is easy to see that operations defined by formulas 
%\begin{eqnarray}
%&& [a,b]_h=(a,b)_h-(-1)^{n_a n_b}(b,a)_h,\nonumber\\
%&& [a,b,c]_h=(a,b,c)_h+(-1)^{(n_a +n_b)n_c}(c,a,b)_h-(-1)^{n_bn_a}(b,a,c)_h-\\
%&&(-1)^{n_cn_b+n_a(n_b+n_c)}(c,b,a)_h+(-1)^{n_a(n_b+n_c)}(b,c,a)_h-
 %(-1)^{n_bn_c}(a,c,b)_h\nonumber
 %\end{eqnarray}
 %satisfy relations of $L_{\infty}$-algebra. 
 
In the next section, we will construct the associative algebra related to the first order Yang-Mills theory. \\
 
\noindent {\bf 3.2. The first order Maxwell complex and the related associative algebra \cite{costello}.} 
Let us consider the following complex: 
\begin{eqnarray}
0\xrightarrow{ }\mK^0\xrightarrow{\t\mQ }\mK^1\xrightarrow{ \t\mQ}\mK^2\xrightarrow{ \t\mQ}\mK^3\xrightarrow{ }0
\end{eqnarray}
such that $\mK^0=\Omega^{0}(M)$, $\mK^{1}=\Omega^{1}(M)\oplus\Omega_+^{2}(M)$, $\mK^{2}=\Omega^{3}(M)\oplus\Omega^{2}_+(M)$, 
$\mK^{3}=\Omega^{4}(M)$ and the differential $\t \mQ$ acts as follows:
\[
\xymatrixcolsep{30pt}
\xymatrixrowsep{3pt}
\xymatrix{
0\ar[r]&\Omega^0(M) \ar[r]^\ud & \Omega^1(M) \ar[ddddr]  &\Omega^3(M) \ar[r]^\ud & \Omega^4(M)\ar[r]&0\\
 &\quad &     & _{-\ud}\quad    &&\\
 && \bigoplus & \bigoplus     &&\\
 &&           & _{\ud_{+}}\quad    &&\\
 && \Omega^2_+(M) \ar[r]_{-Id} \ar[uuuur]  & \Omega^2_+(M)&
}
\]
where $\ud_{+}=\ud+*\ud$ and $\Omega^2_+(M)$ is the space of self-dual 2-forms on the manifold $M$.  
One can define a bilinear operation on the resulting complex:
\begin{center}
$\mu (f_1,f_2)$=
\end{center}
\begin{tabular}{|l|c|c|c|r|}
\hline
\backslashbox{ $f_2$}{$f_1$}&          $v_1$ & $(\mA_1,{\bf F}_1)$ & $(\mW_1,{\bf G}_1)$ & $a_1$ \\
\hline
$v_2$ &                  $v_1v_2$    &$(v_2\mA_1,v_2 {\bf F}_1)$ &$(v_2\mW_1,v_2{\bf G}_1)$ &$v_2a_1$\\
\hline
$(\mA_2,{\bf F}_2)$     &  $(v_1\mA_2,v_1{\bf F}_2)$ & $(\mA_2\wedge {\bf F}_1-$   &   $-\mA_2\wedge \mW_1-$&0\\
                               &                               &    $\mA_1\wedge {\bf F}_2,2P_{+}(\mA_1\wedge\mA_2)$   & ${\bf F}_2\wedge {\bf G}_1$&\\                    
\hline
$(\mW_2,{\bf G}_2)$ & $v_1\mW_2$ &  $-\mA_1\wedge \mW_2-$ &    0&0\\
&&                        ${\bf F}_1\wedge {\bf G}_2$&&\\
\hline
$a_2$                     & $v_1a_2$ &   0 &    0&0\\
\hline
\end{tabular}\\
\vspace{3mm}

\noindent where $P_{+}=\frac{1+*}{2}$ is the projection operator on $\Omega_+^2(M)$. 
Here $f_1,f_2$ take values in the set of variables with prime and  double prime correspondingly. 
Other elements in the table correspond to the appropriate values of $\mu(f_1,f_2)$. In the table above 
$v_1,v_2\in \mK^0$; $(\mA_1,{\bf F}_2), (\mA_2,{\bf F}_2)\in\mK^1=\Omega^1(M)\oplus \Omega^2_+(M)$;  $(\mW_1,{\bf G}_1),(\mW_2,{\bf G}_2)\in\mK^2=\Omega^{3}(M)\oplus \Omega_+^{2}(M)$;  $a_1,a_2\in\mK^3$. It is not hard to see that this operation gives to the complex above the structure of differential graded abelian algebra. If we tensor it with some Lie algebra $\mg$, 
we will find that it possesses the cyclic structure, where the corresponding pairing on $(\mK,\t Q)$ is defined by the same formula as in the previous section. 
One can obtain that it is related to the following action:
\begin{eqnarray}
S=\int_M\Tr\Big({\bf F}\wedge{\bf F}+{\bf F}\wedge (\ud \mA+\mA\wedge\mA)\Big),
\end{eqnarray}
where $F\in\Omega^{+}(M)$, which is equivalent to the usual YM theory. 
Here we note that we could choose the space of anti self-dual 2-forms $\Omega^2_-(M)$ instead of $\Omega^2_+(M)$ in the complex above, and the resulting complex will also have the structure of differential graded abelian algebra.
\\

\noindent {\bf 3.3. Transfer of the $A_{\infty}$ structure.} 
Suppose, $\mF$ is a differential graded algebra (DGA), i.e. it is a chain complex with a differential $Q'$, and $m$ is an associative bilinear operation on $\mK$. Suppose we have a chain complex $\mK$ with a differential $Q'$, which is homotopically equivalent to $\mF$, i.e. there are chain maps $f: \mF\to \mK$ and $g:\mK\to \mF$, such that  $fg-id=Q'H+HQ'$, where $H$ is of degree $-1$. Then $\mF$ and $\mK$ are quasiisomorhic, i.e. we have isomorphism on the level of cohomology: $H_{Q'}^*(\mK)\cong H_{Q}^*(\mF)$ and $\mu_2(\cdot,\cdot)=g\circ m(f(\cdot), f(\cdot))$ is an associative bilinear operation on $H_{Q'}^*(\mK)$. Therefore, we have a $transfer$ of the associative multiplication on the level of cohomology. 
As we explained in the introduction, there is a transfer on the level of complexes, but not of the associative algebra. The structure induced from the operation $m$ and the homotopical equivalence is the $A_{\infty}$-algebra. An elementary calculation shows that 
\begin{eqnarray}
&&\mu_2(\mu_2(a,b),c)-\mu_2(a,\mu_2(b,c))=\\
&&Q\mu_3(a,b,c)+\mu_3(Qa,b,c)+(-1)^{|a|}\mu_3(a,Qb,c)+(-1)^{|a|+|b|}\mu_3(a,b,Qc),\nonumber
\end{eqnarray}
where 
\begin{eqnarray}\label{tran}
&&\mu_3(a,b,c)=\\
&&g\circ m(H\circ m(f(a),f(b)),f(c))-(-1)^{|a|}g\circ m(f(a),H\circ m(f(a),f(b))\nonumber
\end{eqnarray}
is a trilinear operation of degree $-1$. One can find that $\mu_3$ also satisfies bilinear ``higher associativity" relations with $\mu_2$ and $\mu_1\equiv Q'$ and leads to next operation $\mu_4$. One can continue the process and in general obtain the infinite amount of operations satisfying certain billinear relations. The general construction of such $\mu_n$ can be described by 
certain sum over the expressions parametrized by tree graphs with vertices corresponding to operation $m$ and edges corresponding to homotopy $H$ \cite{merk}, \cite{konts}. 
In fact, there is the more general statement: the differential graded algebra structure can be replaced by $A_{\infty}$ structure, and it again will be transferred to the $A_{\infty}$-algebra (see Appendix A and e.g. \cite{markltransf}). 

Now we will show that one can transfer the $A_{\infty}$ structure from the first order complex to the Maxwell one, according to \cite{markltransf}. Firstly, we will show that the Maxwell complex and the first order complex are quasiisomorphic.  
Namely let us construct the maps $g:(\mK^{\cdot}, \t \mQ)\to  (\mF^{\cdot}, \mQ)$, $f:
(\mF^{\cdot}, Q)\to  (\mK^{\cdot}, \t \mQ)$ such that their composition is homotopic to identity. 
The explicit expression for $f$ and $g$ are:
\begin{eqnarray}
&& f(u)=u,\quad f(\mA)=(\mA,2P_{+}\ud \mA), \quad f(\mV)=(\mV,0),\quad f(v)=v,\nonumber\\
&& g(u)=u, \quad g((\mA,{\bf F}))=\mA,\quad g((\mV, \mG))=d\mG+\mV,\quad g(v)=v.
\end{eqnarray}
Here $u\in\Omega^{0}$, $\mA\in\Omega^{1}$, ${\bf F},\mG\in\Omega_{+}^2$, 
$\mV\in \Omega^3$, $v\in\Omega^4$.
Therefore, 
\begin{eqnarray}
g\circ f=id,\quad  f\circ g=id +\t \mQ H +H\t \mQ
\end{eqnarray}
Here $H$ is the homotopy on the complex $(\mK^{\cdot},\t \mQ)$, i.e. the map 
$H: \mK^{i}\to\mK^{i-1}$, and it is nonzero only on $\mK^2$. The explicit formula is:
\begin{eqnarray}
 H((\mV,\mG))=(0,\mG)
\end{eqnarray}
Hence,
\begin{eqnarray}
&& f\circ g((\mA,\mathbf{F}))=(\mA,2P_{+}d\mA), \quad f\circ g((\mV,\mG))=(d\mG+\mV,0),\nonumber\\
&& \t \mQ H((\mV,\mG))=(d\mG,-\mG), \quad H\t \mQ((\mA,\mathbf{F}))=(0,2P_{+}d\mA-\mathbf{F}).
\end{eqnarray}
Therefore we have a Proposition.\\

\noindent{\bf Proposition 3.2.} {\it 
i)The complexes $(\mF^{\cdot},\mQ)$, $(\mK^{\cdot}, \t \mQ)$ are 
quasiisomorphic and homotopically equivalent provided by the maps $f,g$.\\
ii)Under the homotopy equivalence, the structure of DGA (from subsection 2.2) on $(\mK^{\cdot}, \t \mQ)$ is transferred to the described above $A_{\infty}$ structure (from subsection 2.1)on $(\mF^{\cdot},\mQ)$ .}\\

\noindent{\bf Proof.} We have proved $(i)$ above. Let us prove $(ii)$. Namely, consider for example how the 
$A_{\infty}$ structure is transferred from $(\mK^{\cdot},\t \mQ)$ to $(\mF^{\cdot},\mQ)$. 
The bilinear and trilinear operations that are transferred have the following form \rf{tran}:
\begin{eqnarray}
&&(a_1,a_2)'_h=g\circ\mu(f(a_1),f (a_2)), \nonumber\\
&&(a_1,a_2,a_3)'_h=g\circ\mu(f(a_1),H\circ \mu(f(a_2),f(a_3))-\nonumber\\
&&(-1)^{|a_1|}\mu\circ (H\circ\mu(f(a_1),f(a_2)), f(a_3)), 
\end{eqnarray}
where $a_i\in (\mF^{\cdot},\mQ)$ $(i=1,2,3)$. The simple calculation shows that $(a_1,a_2)'_h$, $(a_1,a_2,a_3)'_h$ 
coincide with the defined earlier $(a_1,a_2)_h$, $(a_1,a_2,a_3)_h$. $\blacksquare$\\

\noindent  The example we have considered in this subsection shows that some of the field theory reformulations can be described in terms of homological algebra only. In the next section we will see the same picture in the case of supersymmetric four-dimensional Yang-Mills theory and its first order reformulation.

\section{$A_{\infty}$-algebras of superforms in 4D.}

\noindent{\bf 4.1. SUSY Maxwell complex for complex superfields.} We  claim that the Maxwell complex 

\begin{equation}\label{max}
0\xrightarrow{ }\Omega^{0}(M)\xrightarrow{\ud}\Omega^{1}(M)
\xrightarrow{\m}\Omega^{3}(M)\xrightarrow{\ud}\Omega^{4}(M)\to 0
\end{equation}
and the complex of the first order theory:
\begin{eqnarray}\label{fo}
\xymatrixcolsep{30pt}
\xymatrixrowsep{3pt}
\xymatrix{
0\ar[r]&\Omega^0(M) \ar[r]^\ud & \Omega^1(M) \ar[ddddr]  &\Omega^3(M) \ar[r]^\ud & \Omega^4(M)\ar[r]&0\\
 &\quad &     & _{-\ud}\quad    &&\\
 && \bigoplus & \bigoplus     &&\\
 &&           & _{\ud_{+}}\quad    &&\\
 && \Omega^2_+(M) \ar[r]_{-Id} \ar[uuuur]  & \Omega^2_+(M)&
}
\end{eqnarray}
where $M=\mathbb{R}^{1,3}$, can be naturally embedded into certain complex in N=1 superspace. This is a superspace where in addition to four even spacetime coordinates, there are two odd Weyl spinor coordinates 
$\theta^{\alpha}, \b \theta^{\dot\alpha}$.  
We keep the notation close to \cite{susybook}, see also Appendix B. Denote the space of complex scalar superfields $V(x,\theta,\b \theta)$ as $\Sigma$ and the space of chiral spinor superfields  $W_{\alpha}(x,\theta)$ (corresponding to field strength) as $\Theta$. Finally, we denote the space of chiral superfields corresponding to gauge transformations $\Lambda(x,\theta)$ as $\Phi$. The following complex:
\begin{eqnarray}\label{msusy}
0\xrightarrow{ }\Xi\xrightarrow{id}\Sigma
\xrightarrow{D^{\alpha}{\b D}^2D_{\alpha}}\t \Sigma\xrightarrow{{\b D}^2}\t \Xi\to 0.
\end{eqnarray}
where $\Xi\cong\t \Xi$ and $\Sigma\cong \t\Sigma$ is the supersymmetric generalization of the Maxwell complex, it is clear that the Maxwell complex \rf{max} naturally embeds into the complex \rf{msusy}. 
Moreover, one can consider the complex
\begin{eqnarray}\label{sfo}
\xymatrixcolsep{30pt}
\xymatrixrowsep{3pt}
\xymatrix{
0\ar[r]&\Xi \ar[r]^{id} & \Sigma \ar[ddddr]  &\tilde\Sigma \ar[r]^{\b D^2} & \tilde\Xi\ar[r]&0\\
 &\quad &     & _{\widehat{div}}\quad    &&\\
 && \bigoplus & \bigoplus     &&\\
 &&           & \quad _{\b D^2D_{\alpha}}  &&\\
 && \Theta \ar[r]_{Id} \ar[uuuur]  & \tilde\Theta&
}
\end{eqnarray}
where  
$\widehat{div}{W}=D^{\alpha}W_{\alpha}$, $W$ is a chiral spinor field from 
$\Sigma$ and $\tilde\Theta\cong \Theta$. One can show that \rf{fo} can be embedded in \rf{sfo}. 
As we noted in the section 2.2, one could consider the first order Maxwell complex with antiselfdual 2-forms. This complex will be embedded into antichiral version of the \rf{sfo} (i.e. the spaces $\Xi$ and $\Theta$ would be replaced by their antichiral counterparts). 

Afterwards, we will denote the supersymmetric Maxwell complex as  $(\mFs^{\cdot},\md)$ and the  
supersymmetric complex for the first order theory as $( \mKs^{\cdot}, \tmd)$. 
As one may suspect, these complexes are homotopy equivalent as in the previous case. 
One can construct the supersymmetric version of maps $f,g$ from the previous subsection, i.e. maps $g^s:(\mKs^{\cdot}, \tmd)\to  (\mFs^{\cdot}, \md)$, $f^s: 
(\mFs^{\cdot}, \md)\to  (\mKs^{\cdot}, \tmd)$:
\begin{eqnarray}\label{fsgs}
f^s(\Lambda)=\Lambda, \quad f^s(V)=(-V,\b D^2D_{\alpha}V), \quad f^s(\t V)=(\t V, 0),\quad f^s(\t \Lambda)=\t \Lambda,\\
g^s(\Lambda)=\Lambda, \quad g^s((V,W))=-V, \quad g^s((\t V, \t W_{\alpha})=\t V-D^{\alpha}\t W_{\alpha},\quad g^s(\t \Lambda)=\t \Lambda.\nonumber
\end{eqnarray}
Moreover, $g^s\circ f^s=id$ and $f^s\circ g^s=id +\tmd H+H\tmd$ where the homotopy operator is nonzero only on $\mKs^2$ and is defined by the formula:
\begin{equation}
H((V,W))=((0,-W)).
\end{equation}
The precise statement is the following.\\

\noindent{\bf Proposition 4.1.} {\it The complexes $(\mKs^{\cdot}, \tmd)$, $(\mFs^{\cdot}, \md)$ are quasiisomorphic and homotopically equivalent where the corresponding maps $g^s:(\mKs^{\cdot}, \tmd)\to  (\mFs^{\cdot}, \md)$, $f^s:(\mFs^{\cdot}, Q)\to  (\mKs^{\cdot}, \tmd)$ are given by \rf{fsgs}.}\\

Another issue about two complexes above is that one can define graded antisymmetric bilinear forms on 
$\mF_{susy}$, $\mK_{susy}$ such that 
\begin{equation}
\langle\cdot, \cdot \rangle_{\mK}:  \mK^i_{susy}\otimes  \mK^{3-i}_{susy}\to \mathbb{C},\quad
\langle\cdot, \cdot \rangle_{\mF}:  \mF^i_{susy}\otimes  \mF^{3-i}_{susy}\to \mathbb{C}.
\end{equation}
On $\mK_{susy}$ they are defined by the following formulas:
\begin{eqnarray}
&&\langle{\Lambda,\t\Lambda}\rangle_{\mK}=\int d^4 xd^2\theta \Lambda\t \Lambda,\quad \langle V,\t V\rangle{\mK}=\int d^4 x d^4\theta V\t V,\nonumber\\
&& \langle W,\t W\rangle_{\mK}=\frac{1}{2}\int d^4 xd^2\theta W_{\alpha}\t W^{\alpha}.
\end{eqnarray}

\noindent {\bf Proposition 4.2.} {\it The graded symmetric bilinear forms $\langle\cdot, \cdot \rangle_{\mF}$, $\langle\cdot, \cdot \rangle_{\mK}$ obey the relation 
\begin{eqnarray}\label{cycpairing}
\langle a_1, Qa_2\rangle+(-1)^{a_1+a_2+a_1a_2}\langle a_2,Qa_1\rangle=0, 
\end{eqnarray}
when $a_1,a_2\in \mF^r$ or $a_1,a_2\in \mK^r$ and $Q$ stands for $\md$ or $\tmd$.}\\

\noindent The bilinear form on $\mF_{susy}$ is defined by the restriction of the form on $\mK_{susy}$. Therefore, from now on we suppress the index notation for the bilinear forms.

One of the immediate consequences of the Proposition above is that if one can construct the $A_{\infty}$-algebra on the complex $(\mKs^{\cdot}, \tmd)$, then the $A_{\infty}$ structure on 
$(\mFs^{\cdot}, \md)$ can be obtained by the transfer procedure as in the previous section. \\

\noindent {\bf 4.2. Reminder of N=1 SUSY YM.} Let us consider the action of the N=1 SUSY YM theory in 4 dimensions:
\begin{eqnarray}\label{asusy}
S_{SYM}[V]=\frac{1}{2}\int d^4xd^4 \theta ~{\rm Tr} (e^{-V}D_{\alpha}e^V{\b D}^2(e^{-V}D_{\alpha}e^V)),
\end{eqnarray}
where $V$ is the real superfield, taking values in some Lie algebra $\mathfrak{g}$, i.e. $V\in\Sigma\otimes{\mathfrak{g}}$. It is invariant under the following transformations:
\begin{eqnarray}
\label{transf}
e^V\to e^{\bar\Lambda}e^{V}e^{\Lambda},
\end{eqnarray}
where $\Lambda$ is a chiral scalar superfield. 
%When $\Lambda, \b \Lambda$ is infinitesimal it is: 
%\begin{eqnarray}
%\delta (V)=\frac{1}{2}L_V(\Lambda-\b \Lambda+coth(\frac{1}{2}L_V)(\Lambda+\b \Lambda)).
%\end{eqnarray}
%Sometimes it is useful to define the following representation of the field V:
%\begin{eqnarray}\label{chiral}
%e^V=e^{\b \Omega}e^{\Omega},
%\end{eqnarray}
%such that under \rf{transf}, the transformation of $\Omega$ and $\bar{\Omega}$ is:
%\begin{eqnarray}
%e^{\Omega}\to e^{\Omega}e^{\Lambda}, \quad e^{\b \Omega}\to e^{\b \Lambda}e^{\b \Omega}.
%\end{eqnarray}
It is useful to define the supercurvature, i.e. the spinor chiral superfield $W\in \Theta\otimes{\mathfrak{g}}$
\begin{eqnarray}
W_{\alpha}=-\b D^2(e^{-V}D_{\alpha}e^V).
\end{eqnarray} 
Using the global transformation formula \rf{transf}, one can find the infinitesimal transformations of $V$ and $W$:  
\begin{eqnarray}
&&V\to V+\delta_{\Lambda, \b \Lambda}V=V+\frac{1}{2}L_{V}(\Lambda-\bar \Lambda+\coth(\frac{1}{2}L_V)(\Lambda+\b \Lambda))\nonumber\\
&&W_{\alpha}\to W_{\alpha}+[W_{\alpha},\Lambda]
\end{eqnarray} 
where $\Lambda, \bar{\Lambda}$ are infinitesimal and $L_V\cdot=[V,\cdot]$.
Varying the action $S_{SYM}$ with respect to $V$, one finds the following expression:
\begin{eqnarray}\label{eqm}
f(L_V)[\nabla_{\alpha},W^{\alpha}]=0, 
\end{eqnarray}
where $\nabla_{\alpha}=e^{-V}D_{\alpha}e^V$, and $f(x)=\frac{e^{x}-1}{x}$. 
One can see that this equation is equivalent to the "physically covariant" equation $[\nabla_{\alpha},W^{\alpha}]=0$ because operator $f(L_V)$ is invertible. 
Here we note (this is important for the next subsection) that the equation \rf{eqm} is the equation of motion for the functional \rf{asusy} even when $V$ is not real.  The reality condition gives the constraint:
\begin{eqnarray}
[\nabla_{\alpha},W^{\alpha}]=[\nabla_{\dot{\alpha}},W^{\dot{\alpha}}].
\end{eqnarray}
Below we are showing that the equation \rf{eqm} is covariant from the homological point of view, i.e. we 
are observing that it arises as GMC equation for the certain cyclic $A_{\infty}$-algebra. \\

\noindent {\bf 4.3. Chiral $A_{\infty}$-algebra.} 
One can also rewrite the action \rf{asusy} in the first order form the way we we did in the case of usual Yang-Mills: 
\begin{eqnarray}
&&S_{fo}[V,W]=\frac{1}{2}\int d^4x d^2 \theta \frac{1}{2}{\rm Tr}(W_{\alpha}W^{\alpha}+\bar{D}^2 (e^{-V}D_{\alpha}e^V)W^{\alpha}).
\end{eqnarray} 
We see that both actions $S_{SYM}$ and $S_{fo}$ in the case of abelian Lie algebra 
can be written as $\frac{1}{2}\langle \psi,Q\psi\rangle$, where $\psi$ is an element of degree 
$1$ and $Q$ stands either for $\md$ or $\tmd$.
%Now let us consider the representation \rf{chiral} and the case when $\b \Omega=0$, i.e. $V=\Omega$. Therefore we have only $\Lambda$-symmetry left. 
Let us forget about $\b \Lambda$-symmetry in this section and consider only the one, generated by chiral scalar superfields $\Lambda$.  
Then the BV generalization of the last action looks as follows:
\begin{eqnarray}
&& S^{BV}_{fo}=S_{fo}+\nonumber\\
&&\int d^4xd^4\theta {\rm Tr}(\delta_{C}(V)V^*)+ \int d^4x d^2\theta {\rm Tr}([W^{\alpha},C]W^*_{\alpha}+\frac{1}{2}[C,C]C^*).
\end{eqnarray} 
Here, as usual, ghosts and antifields are $c^{+}\in\Phi[1], V^*\in\t \Sigma[-1]$ and is a real superfield, $W^*\in\t \Sigma[-1]$, $c\in\t \Theta[-2]$. In the actions above $\delta_{c}(V)=\frac{1}{2}L_{V}(c+\coth(\frac{1}{2}L_V)(c))$. 
It is feasible to check that these actions satisfy the Master equation. 
Therefore, an odd vector field ${\bf Q}$ 
that is defined by means of the formula ${\bf Q}\cdot=(S_{fo}^{BV},\cdot)_{BV}$ acts on the fields 
in the following way:
\begin{eqnarray}\label{bfq}
&&{\bf Q}c=\frac{1}{2}[C,C],\nonumber\\
&&{\bf Q}V=\delta_{C}(V),\nonumber\\
&&{\bf Q}V^*=f(L_V)[e^{-V}D_{\alpha}e^V,W^{\alpha}]+\frac{\delta}{\delta V}\delta_{C}(V)V^*,\nonumber\\
&&{\bf Q}W_{\alpha}=[W_{\alpha},C],\nonumber\\
&&{\bf Q}W_{\alpha}^*=[W_{\alpha}^*,C],\nonumber\\
&&{\bf Q}{C^*}=[C,C^*]-\frac{1}{2}L_V(V^*-\coth(\frac{1}{2}L_V)(V^*)) +[W_{\alpha},{W^{\alpha}}^*].
\end{eqnarray}
Then we have the following Proposition.\\

\noindent{\bf Proposition 4.3.} {\it The odd vector field ${\bf Q}$ satisfies the nilpotency condition ${\bf Q}^2=0$ and therefore determines the $A_{\infty}$ structure on the complex \rf{fo}.}\\

\noindent Therefore, the action $S^{BV}_{fo}$  has the form of the homotopy Chern-Simons theory, 
i.e.:
\begin{eqnarray}
\frac{1}{2}\langle\psi,\tmd\psi\rangle+\sum_{n\ge 2}\frac{1}{n+1}\langle\mu_n(\psi,...,\psi),\psi\rangle,
\end{eqnarray}
where $\mu_n$ stand for bilinear and higher operations in the $A_{\infty}$-algebra. 
One of the specific features of the corresponding $A_{\infty}$-algebra is that the 
corresponding $L_{\infty}$-algebra vanishes in the case of abelian Lie algebra 
$\mathfrak{g}$. One can see that directly from \rf{bfq} or from the fact that the BV action in this case becomes bilinear in fields. Therefore, for example, the bilinear operation for this  
$A_{\infty}$-algebra in the case of abelian $\mathfrak{g}$ is commutative, like in the usual Yang-Mills case.  
Let us write the expression for this bilinear operation explicitly.

\begin{center}
$\mu_2(f_1,f_2)$=
\end{center}
\begin{tabular}{|l|c|c|c|r|}
\hline
\backslashbox{$ f_2$}{$f_1$}&          $\Lambda_1$ & $(V_1,W_1)$ & $(\t V_1,\t W_1)$ & $\t\Lambda_1$ \\
\hline
$\Lambda_2$ &                  $\Lambda_1\Lambda_2$    &$(\frac{1}{2}V_1\Lambda_2,W_1\Lambda_2)$ & 
$(\frac{1}{2}\t V_1\Lambda_2,\t W_1\Lambda_2)$ &$\t\Lambda_1\Lambda_2$\\
\hline
$(V_2,W_2)$     &  $(\frac{1}{2}\Lambda_1V_2,\Lambda_1W_2)$ & $((V_1,W_1), (V_2,W_2))_h$   &   $\t W_{1,\alpha}W_2^{\alpha}$ &0\\
                               &                               &       & $-\frac{1}{2}\b D^2(\t V_1V_2)$&\\                    
\hline
$(\t V_2,\t W_2)$ & $(\frac{1}{2}\Lambda_1\t V_2,\Lambda_1\t W_2)$ &  $-W_1^{\alpha}\t W_{2,\alpha}$ &    0&0\\
&&                        $-\frac{1}{2}\b D^2(V_1\t V_2)$&&\\
\hline
$\t\Lambda_2$                     & $\Lambda_1\t\Lambda_2$ &   0 &    0&0\\
\hline
\end{tabular}\\
\vspace{3mm}

Here $f_1,f_2$ take values in the set of variables with indices $1$ and $2$ correspondingly. In the table above $\Lambda_i\in \mKs^0$, $(V_i,W_i)\in \mKs^1 (i=1,2)$, $(\t V_i,\t W_i)\in \mKs^2 (i=1,2)$, 
$\t \Lambda_i\in \mKs^3$, and 
\begin{eqnarray}
&&((V_1,W_1), (V_2,W_2))_h=\\
&&(D_{\alpha}V_1W_2^{\alpha}+W_1^{\alpha}D_{\alpha}V_2+\frac{1}{2}(V_1D_{\alpha}W^{\alpha}_2-D_{\alpha}W^{\alpha}_1V_2),-\frac{1}{2}\b D^2(V_1D_{\alpha}V_2-D_{\alpha}V_1V_2).\nonumber
\end{eqnarray}
{\bf Corollary 4.1.} {\it The operation $\mu$ on the complex \rf{sfo} is homotopy associative on 
$(\mathfrak{K}^{\cdot}, \tilde{\mathfrak{D}})$}.\\

\noindent In Appendix B, we explicitly prove this proposition.

Knowing all the operations of $A_{\infty}$-algebra based on complex \rf{sfo}, one can construct ones on the complex \rf{msusy}. The resulting bilinear operation is:
\begin{center}
$m_2(f_1,f_2)$=
\end{center}
\begin{tabular}{|l|c|c|c|r|}
\hline
\backslashbox{$ f_2$}{$f_1$}&          $\Lambda_1$ & $V_1$ & $\t V_1$ & $\t\Lambda_1$ \\
\hline
$\Lambda_2$ &                  $\Lambda_1\Lambda_2$    &$\frac{1}{2}V_1\Lambda_2$ & 
$\frac{1}{2}\t V_1\Lambda_2$ &$\t\Lambda_1\Lambda_2$\\
\hline
$V_2$     &  $\frac{1}{2}\Lambda_1V_2 $ & $(V_1,V_2)_h$   &  $-\frac{1}{2}\b D^2(\t V_1V_2)$  &0\\
                    
\hline
$\t V_2$ & $\frac{1}{2}\Lambda_1\t V_2$ &  $-\frac{1}{2}\b D^2(V_1\t V_2)$ &    0&0\\
\hline
$\t\Lambda_2$                     & $\Lambda_1\t\Lambda_2$ &   0 &    0&0\\
\hline
\end{tabular}\\

\noindent where 
\begin{eqnarray}
&& (V_1,V_2)_h=\frac{1}{2}D^{\alpha}\b D^2(V_1D_{\alpha}V_2-V_2D_{\alpha}V_1)-D_{\alpha}V_1\b D^2D^{\alpha}V_2-\nonumber\\
&&D_{\alpha}V_2\b D^2D^{\alpha}V_1D_{\alpha}V_2+\frac{1}{2}(V_2D_{\alpha}\b D^2D^{\alpha}V_1-V_1D_{\alpha}\b D^2D^{\alpha}V_2).
\end{eqnarray}

\noindent Therefore, we have another proposition.\\

\noindent {\bf Proposition 4.4.} {\it Operation $m_2$ is homotopy associative on the supersymmetric generalization of the Maxwell complex \rf{msusy}.}\\

\noindent As one could expect, the operation $m_2$ comes from the BV functional
 \begin{eqnarray}\label{bvsymact}
&&S^{BV}_{SYM}[V]=\frac{1}{2}\int d^4xd^4 \theta ~{\rm Tr}(e^{-V}D_{\alpha}e^V{\b D}^2(e^{-V}D_{\alpha}e^V))+\nonumber\\
&&\int d^4xd^4\theta ~{\rm Tr}(\delta_{C}(V)V^*) +\int d^4xd^2\theta\frac{1}{2}{\rm Tr}([C,C]C^*).
\end{eqnarray}

Again, as in the case of nonsupersymmetric Yang-Mills, we see that the $A_{\infty}$-algebras of the second and the first order theories are related by means of transfer formula. \\

\noindent {\bf 4.4. Full symmetry and real representation.} So far we neglected the full symmetry \rf{transf} of the action \rf{asusy}. Namely, 
we considered only the chiral part of the symmetry. In order to obtain full symmetry on the homological level there are two options. The first one is to consider the following extension of the complex \rf{msusy}:
\begin{eqnarray}
\xymatrixcolsep{40pt}
\xymatrixrowsep{3pt}
\xymatrix{
0\ar[r]&\Theta \ar[r]^{id} & \Sigma \ar[r]^{D_{\alpha}\b D^2D^{\alpha}}  &\tilde\Sigma \ar[r]^{\b D^2}\ar[ddr]^{D^2} & \tilde\Theta\ar[r]&0\\
 &\oplus&&&\oplus
 &&          &  &&\\
 &\bar{\Theta}\ar[uur]^{-id}&   &&\tilde{\bar{\Theta}}}\nonumber
\end{eqnarray}
Here $\bar{\Theta}\cong \bar{\tilde \Theta}$ is a space of antichiral superfields. From now on, we denote it as $(\mF_{full}^{\cdot}, \mathcal{D})$. 
 To reproduce this complex, one should consider the full BV action:
\begin{eqnarray}
&&S^{BV, full}_{SYM}[V]=\nonumber\\
&&\frac{1}{2}\int d^4x d^4 \theta ~{\rm Tr}(e^{-V}D_{\alpha}e^V{\b D}^2(e^{-V}D_{\alpha}e^V))+\int d^4xd^4\theta ~{\rm Tr}(\delta_{C,\b C}(V)V^*)+\nonumber\\
&&\int d^4x d^2\theta\frac{1}{2}{\rm Tr}([C,C]C^{*})+\int d^4x d^2\b \theta\frac{1}{2}{\rm Tr}([\b C,\b C]\b C^*).
\end{eqnarray}
Here $\delta_{C,\b C}(V)=\frac{1}{2}L_V(C-\b C+coth(\frac{1}{2}L_V)(C+\b C))$. The corresponding 
bilinear operation should be modified in this way:
\begin{center}
$m^{full}_2(f_1,f_2)$=
\end{center}
\begin{tabular}{|l|c|c|c|r|}
\hline
\backslashbox{$ f_2$}{$f_1$}&          $(\Lambda_1, \b\Lambda_1)$ & $V_1$ & $\t V_1$ & $(\t\Lambda_1, 
\b{\t{\Lambda}}_1)$ \\
\hline
$(\Lambda_2, \b \Lambda_2)$ &                  $(\Lambda_1\Lambda_2, \b \Lambda_1\b\Lambda_2)$    &$\frac{1}{2}V_1(\Lambda_2+\b\Lambda_2)$ & 
$\frac{1}{2}\t V_1(\Lambda_2+\b\Lambda_2)$ &$(\t\Lambda_1\Lambda_2,\b{\t{\Lambda}}_1\b{\t{\Lambda}}_2)$\\
\hline
$V_2$     &  $\frac{1}{2}(\Lambda_1+\b\Lambda_1)V_2 $ & $(V_1,V_2)_h$   &  $(-\frac{1}{2}\b D^2(\t V_1V_2), \frac{1}{2}D^2(\t V_1V_2))$  &0\\
                    
\hline
$\t V_2$ & $\frac{1}{2}(\Lambda_1+\b\Lambda_1)\t V_2$ &  $-\frac{1}{2}\b D^2(V_1\t V_2)$ &    0&0\\
\hline
$(\t\Lambda_2, {\b{\t\Lambda}}_2)$                     & $(\Lambda_1\t\Lambda_2, \b\Lambda_1{\b{\t\Lambda}}_2)$ &   0 &    0&0\\
\hline
\end{tabular}\\

\noindent Using the transfer formula, one can find the modification for the bilinear operation of the first order model complex. 

Sometimes it is useful to define the following representation of the field $V$:
\begin{eqnarray}\label{chiral}
e^V=e^{\b \Omega}e^{\Omega},
\end{eqnarray}
such that under \rf{transf}, the transformation of $\Omega$ and $\bar{\Omega}$ is:
\begin{eqnarray}
e^{\Omega}\to e^Ke^{\Omega}e^{\Lambda}, \quad e^{\b \Omega}\to e^{\b \Lambda}e^{\b \Omega}e^{-K},
\end{eqnarray}
where the new $K$-symmetry appears. 
Therefore, another approach to get full symmetry into the picture is to represent $V$ in terms of $\Omega$ and $\b \Omega$ and treat them as separate varibles in the BV action. The resulting $A_{\infty}$-algebra is based on the following complex:
\begin{eqnarray}
\xymatrixcolsep{40pt}
\xymatrixrowsep{3pt}
\xymatrix{
&\Theta \ar[r]^{id} & \Sigma_c\ar[ddddr] \ar[r]^{D_{\alpha}\b D^2D^{\alpha}}  &\tilde\Sigma_c \ar[ddr]^{id}\ar[r]^{\b D^2} & \tilde\Theta\\
 &\oplus&& &\oplus&&          &  &&\\
 0\ar[r]&\Upsilon\ar[uur]^{id} \ar[ddr]^{-id}&\oplus&\oplus&\t \Upsilon\ar[r]&0\\
 &\oplus&&&\oplus
 &&          &  &&\\
 &\bar{\Theta}\ar[r]^{-id}&\bar\Sigma_c\ar[uuuur]  \ar[r]^{D_{\alpha}\b D^2D^{\alpha}} &\bar{\tilde\Sigma}_c\ar[uur]^{-id}\ar[r]^{D^2}&\tilde{\bar{\Theta}}}\nonumber
\end{eqnarray}
Here all maps in the middle of the diagram are given by the action of the operator $D_{\alpha}\b D^2D^{\alpha}$, and the spaces $\Sigma_r, \tilde\Sigma_r, \bar\Sigma_r, \bar{\tilde\Sigma}_r,\Upsilon$ are isomorphic to the space $\Sigma$ (see above). 
We will denote the resulting complex as $(\mF^{\cdot}_{real}, \mathcal{D})$. The immediately arising interesting question is: how the $A_{\infty}$ structure on this complex is related to $A_{\infty}$-algebra on 
$(\mF_{full}^{\cdot}, \mathcal{D})$? Below  we will show that they are homotopically equivalent. 
Let us explicitly construct the chain maps $f:\mF_{full}^{\cdot}\to \mF_{real}^{\cdot}$ and $g:\mF_{real}^{\cdot}\to \mF_{full}^{\cdot}$:
\begin{eqnarray}
&&f((\Lambda, \b \Lambda))=(\Lambda, -\frac{1}{2}(\Lambda+\b \Lambda), \b \Lambda),\quad 
f(V)=(\frac{1}{2}V, \frac{1}{2}V), \nonumber\\
&& f(\t V)=(\t V, \t V), \quad f((\Lambda, \b \Lambda))=(\Lambda, 0, \b \Lambda)\nonumber\\
&&\nonumber\\
&&g((\Lambda, K, \b \Lambda))=(\Lambda, \b \Lambda), \quad g((\Omega, \b \Omega))=\Omega+\b \Omega\nonumber\\
&&g((\t \Omega, \t{\b\Omega}))=\frac{1}{2}(\t \Omega+ \t {\b\Omega}), \quad g((\Lambda, K, \b \Lambda))=
(\Lambda-\frac{1}{2}{\b D}^2K, \b \Lambda-\frac{1}{2}{D}^2K)
\end{eqnarray}
It is feasible to check that $gf=id$ and $fg=id+[Q,H]$ where $H$ is an operator of degree $-1$ on the complex 
$\mF_{chiral}^{\cdot}$ which is nonzero on $\mF^1_{real}$ and $\mF^3_{real}$, such that 
\begin{eqnarray}
&&H((\Omega, \b \Omega))=(0, \frac{1}{2}(\b \Omega-\Omega),0),\nonumber\\
&&H((\t \Lambda, \t K, \t {\b \Lambda}))=(-\frac{1}{2}K,\frac{1}{2}K) 
\end{eqnarray}
Therefore, one can formulate the following Proposition.\\

\noindent {\bf Proposition 4.5.} {\it Complex $(\mF_{full}^{\cdot}, \mathcal{D})$ is homotopically equivalent to 
$\mF_{real}^{\cdot}$.}\\

This allows us to transfer the homotopy algebra structure from $\mF_{real}^{\cdot}$ to $\mF_{full}^{\cdot}$. Here we also note the following. If we look at  "Baker-Campbell-Hausdorff" change of variables 
$e^V=e^{\b\Omega}e^{\Omega}$ in the corresponding equation of motion (Maurer-Cartan equation), we find out that it becomes part of the transfer of the $A_{\infty}$-algebra.

\section*{Appendix A: $A_{\infty}$-algebras and the BV formalism.}
In this appendix we summarize all necessary information about $A_{\infty}$-algebras. For more details see 
e.g. \cite{stashbook}, \cite{kajiura}, \cite{markltransf}. \\

\noindent {\bf A1. $A_{\infty}$-algebras.} 
The $A_{\infty}$-algebra is a generalization of an associative algebra with a differential. Namely, consider a graded vector space $V=\oplus_k V_k$ with a differential $Q$. Consider the multilinear operations $\mu_r: V^{\otimes r}\to V$ of the degree $2-r$, such that $\mu_1=Q$. \\

\noindent {\bf Definition.}{\it The space V is an $A_{\infty}$-algebra if $\mu_n$ satisfy the following bilinear identity:
\begin{eqnarray}\label{arel}
\sum^{n-1}_{i=1}(-1)^{i}M_i\circ M_{n-i+1}=0
%(-1)^{k+\lambda+k\lambda+nk+k(n_{a_1}+...+n_{a_{\lambda}})}\mu_{n-k+1}
%(a_1, . . . , a_\lambda,
%\mu_k(a_{\lambda+1}, . . . , a_{\lambda+k}), a_{\lambda+k+1}, . . . , a_n) = 0,
\end{eqnarray}
on $V^{\otimes n}$. 
Here $M_s$ acts on $V^{\otimes m}$ ($m\ge s$)  as the sum of all possible operators of the form 
${\bf 1}^{\otimes^l}\otimes\mu_s\otimes{\bf 1}^{\otimes^{m-s-l}}$ taken with appropriate signs. In other words, 
$M_s:V^{\otimes^m}\to V^{\otimes^{m-s+1}}$ and 
\begin{eqnarray}
M_s=\sum^{m-s}_{l=0}(-1)^{l(s+1)}{\bf 1}^{\otimes^l}\otimes\mu_s\otimes{\bf 1}^{\otimes^{m-s-l}}.
\end{eqnarray}}
Let's write several relations which are satisfied by $Q$, $\mu_1$, $\mu_2$, $\mu_3$:
\begin{eqnarray}
&&Q^2=0,\\
&&Q\mu_2(a_1,a_2)=\mu_2(Q a_1,a_2)+(-1)^{n_{a_1}}\mu_2(a_1,Q a_2),\nonumber\\
&&Q\mu_3(a_1,a_2, a_3)+\mu_3(Q a_1,a_2, a_3)+(-1)^{n_{a_1}}\mu_3(a_1,Q a_2, a_3)_h+\nonumber\\
&&(-1)^{n_{a_1}+n_{a_2}}\mu_3( a_1, a_2, Q a_3)=\mu_2(\mu_2(a_1,a_2),a_3)-\mu_2(a_1,\mu_2(a_2,a_3)).\nonumber
\end{eqnarray}
In such a way we see that if $\mu_n=0$, $n\ge 3$ , then we have just graded differential associative algebra. 
It appears that the relations \rf{arel} can be encoded into one equation $\p^2=0$. To see this, one applies the desuspension operation (the operation which shifts the grading $s^{-1}: V_{k}\to V_{k-1}$) to $\mu_n$, i.e. one can define operations of degree $1$: $\nu_n=s\nu_n {s^{-1}}^{\otimes^n}$. 
More explicitly, 
\begin{eqnarray}
\nu_n(s^{-1} a_1,...,s^{-1} a_n)=(-1)^{s(a)}s^{-1}\mu_n(a_1,...,a_n),
\end{eqnarray}
such that $s(a)=(1-n)n_{a_1}+(2-n)n_{a_2}+...+a_{n-1}$. 
The relations between $\nu_n$ operations can be summarized in the following simple equations:
\begin{eqnarray}\label{M2}
\sum^n_{i=1}N_i\circ N_{n+1-i}=0.
\end{eqnarray}
on $V^{\otimes n}$. Here each $N_s$ acts on $V^{\otimes^m}$ ($m\ge s$) as the sum of all operators ${\bf 1}^{\otimes^l}\otimes\nu_s\otimes{\bf 1}^{\otimes^k}$, such that $l+s+k=m$.  Combining them into one operator $\p=\sum_n\nu_n$, acting on a space $\oplus_kV^{\otimes^k}$ the relations \rf{arel} can be combined into one equation $\p^2=0$. 

Another way to represent the relations \rf{arel} as the nilpotency condition of some operator is by using the differential operators on a noncommutative manifold. Let the set $\{e_i\}$ be the homogeneous elements,  
which form a basis of $V$. Introduce noncommutative supercoordinates $x^i$ such that $X=x^ie_i$ has degree degree 1 on $V$. One can write a noncommutative vector field, such that  
\begin{eqnarray}\label{Q}
{\bf Q}x^i=\nu_{n;j_1,...,j_n}x^{j_1}...x^{j_n},
\end{eqnarray}
where $\nu_{n;j_1,...,j_n}=\nu_n(e_{j_1},...,e_{j_n})$. Then the relations \rf{M2} can be formulated as 
${\bf Q}^2=0$. \\

\noindent{\bf A2. Cyclic structures, BV formalism and the generalized Maurer-Cartan equation.} 
First we define what cyclic structure is.\\

\noindent{\bf Definition.} {\it The $A_{\infty}$-algebra on a space V is called $cyclic$ if there exists a nondegenenerate  pairing $\langle\cdot, \cdot \rangle$, such that it is graded symmetric
\begin{eqnarray}
\langle a, b \rangle=-(-1)^{(n_a+1)(n_b+1)}\langle b,a  \rangle
\end{eqnarray}
and satisfies the following conditions for $\mu_n$:
\begin{eqnarray}
&&\langle a_1,\mu_{n-1}(a_2,...,a_n)\rangle=\nonumber\\
&&(-1)^{(n-1)(a_1+a_2+1)+a_1(a_2+...+a_n)}\langle a_2,\mu_{n-1}(a_3,...,a_n,a_1)\rangle.
\end{eqnarray}}
It makes sense to define $\psi_n(a_1,...,a_n)=\langle a_1, \mu_{n-1}(a_2, ..., a_n)\rangle$. Consider again the element of degree 1 $X=x^ie_i$, where $x^i$ are noncommutative 
supercoordinates and define the formal $action$ $functional$:
\begin{eqnarray}\label{hcs}
S[X]=\sum^{\infty}_{n=2}\frac{1}{n}\Psi_n(X,...,X).
\end{eqnarray}
Let $\omega_{ij}=\langle e_i, e_j\rangle$. Then one can define a BV bracket:
\begin{eqnarray}
(\alpha,\beta)_{BV}=\frac{{\overleftarrow{\p}}\alpha}{\p x^i}\omega^{ij}\frac{\overrightarrow{\p}\beta}{\p x^j}
\end{eqnarray}
on the space of polynomials of $x^i$, such that $S$ satisfies classical Master equation:
\begin{eqnarray}
(S,S)_{BV}=0
\end{eqnarray}
Using this condition, one can find that $S$ defines an odd vector field such that
\begin{eqnarray}
(S,x^i)_{BV}=\nu_{n;j_1,...,j_n}x^{j_1}...x^{j_n},
\end{eqnarray}
which coincides with odd vector field ${\bf Q}$ was defined in \rf{Q}. Moreover, for a given action 
\begin{eqnarray}
S=v_{i_1i_2}x^{i_1}x^{i_2}+\sum_{n\ge3}v_{i_1...i_n}x^{i_1}....x^{i_n},
\end{eqnarray}
which is cyclic in $x^i$ and satisfies Master equation, we obtain the cyclic $A_{\infty}$-algebra (see e.g. \cite{kajiura},\cite{zwiebach}).

Varying the action \rf{hcs} with respect to $X$, one finds the equation of motion which are known as generalized Maurer-Cartan equation:
\begin{eqnarray}
QX+\sum_{n\ge2}\mu_n(X,...,X)=0.
\end{eqnarray}
This equation is known to have the following symmetry
\begin{eqnarray}
X\to Q\alpha+\sum_{n\ge2,k}(-1)^{n-k}\mu_n(X,...,\alpha,...,X),
\end{eqnarray}
where $\alpha$ is an element of degree 0 and $k$ means the position of $\alpha$ in $\mu_n$. \\

\noindent {\bf A3. Transfer of the $A_{\infty}$ structure.} Suppose you have two complexes which are quasiisomorphic and moreover homotopically equivalent. Moreover, suppose that on one of them there exists the structure of $A_{\infty}$-algebra. Then there exists an $A_{\infty}$-algebra on another complex. The explicit formulas are given in \cite{markltransf} following the results of \cite{konts} and \cite{merk}.  

In fact, in \cite{markltransf} there is even a more general statement. 
Let us formulate it in a precise way. Consider two complexes $(\mF,Q)$ and $(\mK, Q')$ such that there are maps 
$f:(\mF,Q)\to (\mK, Q')$, $g:(\mK,Q')\to (\mF,Q)$ such that $fg$ is chain homotopic to identity. In other words, there exists a map $H: (\mK,Q')\to (\mK, Q')$ of degree ${-1}$, such that $fg=id+Q' H+H Q'$. Then, given an $A_{\infty}$-algebra structure $\hat{\mu}_{\{n\}}$ on $\mK$, such that $\hat{\mu}_1=Q'$, one can construct an $A_{\infty}$-algebra on $(\mF,Q)$ by means 
of the formula 
\begin{eqnarray}
\mu_n=g\circ p_n \circ f^{\otimes^n},
\end{eqnarray}
where $p_n$ is obtained by means of consequtive recurrent procedure of applying the homotopy operators to $\t \mu_n$. The explicit formula is:
\begin{eqnarray}
p_n=\sum_B(-1)^{\theta(r_1,...,r_k)}\hat{\mu}_k(H\circ p_{r_1},...,  H\circ p_{r_k}),
\end{eqnarray}
where $B=\{k, r_1, . . . , r_k | 2 \le k \le n, r_1, ... , r_k \ge 1, r_1 + ... + r_k = n\}$, $\theta(r_1,...,r_k)=\sum_{1\le\alpha\le\beta\le_r}r_{\alpha}(r_{\beta}+1)$ and $p_2\equiv \hat{\mu}_2$, $p_1\circ H\equiv id$.

\section*{Appendix B: Explicit calculations for $A_{\infty}$ superalgebras of superforms.}
\noindent {\bf B1.} {\bf Notations for $N=1$ 4D superspace.} We keep the notations from \cite{susybook}. 
The world-volume metric in the 4D space with the coordinates $x^{\mu}$  is $\eta_{\mu\nu}=\rm{diag}(-1,+1,+1,+1)$. 
The $N=1$ four-dimensional space has two anticommutative Weyl spinor coordinates $\theta^{\alpha}, \theta^{\dot{\alpha}}$. The spinor indices are raised and lowered by means of $C_{\alpha\beta}, C_{\dot{\alpha}\dot{\beta}}$, such that $C_{21}=-C_{12}=i$. One can define superderivatives: 
\begin{eqnarray}
D_{\alpha}=\p_{\alpha}+\frac{i}{2}\b \theta^{\dot{\beta}}\p_{\alpha\dot{\beta}}, \quad \b D_{\dot{\alpha}}=\b \p_{\dot{\alpha}}+\frac{i}{2}{\theta}^{\beta}\p_{\beta\dot{\alpha}},
\end{eqnarray}
where $\p_{\alpha\dot{\beta}}=\sigma_{\alpha\dot{\beta}}^{\mu}\p_{\mu}$. 
Therefore, one can introduce the (anti)chiral scalar superfields, which are defined by the nilpotency of certain antiderivatives $D_{\dot{\alpha}}\Lambda(x,\theta)=0$ ($D_{\alpha}\b \Lambda(x,\theta)=0$)
The relations 
between superderivatives are:
\begin{eqnarray}
\{D_{\alpha},D_{\dot{\beta}}\}=i\p_{\alpha\dot{\beta}}, \quad 
\{D_{\alpha},D_{{\beta}}\}=0, \quad \{\b D_{\dot{\alpha}},\b D_{\dot{\beta}}\}=0.
\end{eqnarray}
For calculations it is also useful to introduce operators $D^2=C^{\alpha\beta}D_{\alpha}D_{\beta}$, $\b D^2=C^{\dot{\alpha}\dot{\beta}}D_{\dot{\alpha}}D_{\dot{\beta}}$.\\

\noindent {\bf B2.} {\bf Explicit calculations for the $A_{\infty}$-algebra of $N=1$ SUSY Yang-Mills theory.}  
In this subsection we show by explicit calculation that the bilinear operation we defined on the complex \rf{fo} is homotopy associative and the differential acts on it as a derivation.

For simplicity we denote $\mu_n(\cdot, ..., \cdot)$ as $(\cdot, ..., \cdot)_h$. 
We remind that  we defined the graded symmetric bilinear operation $(\cdot,\cdot)_h$, i.e. 
$(a_1,a_2)_h=(-1)^{|a_1||a_2|}(a_2,a_1)_h$ on the complex \rf{sfo} in the following way:
\begin{eqnarray}
&&(\Lambda_1,\Lambda_2)_h=\Lambda_1\Lambda_2,\nonumber\\
&&(\Lambda,V)_h=\frac{1}{2}\Lambda V,\nonumber\\
&&(\Lambda,W_{\alpha})_h=\Lambda W_{\alpha},\nonumber\\
&&(\Lambda,\tilde W_{\alpha})_h=\Lambda\tilde W_{\alpha},\nonumber\\
&&(\Lambda,\tilde V)_h=\frac{1}{2}\Lambda\tilde V,\nonumber\\
&&(\Lambda,\tilde\Lambda)_h=\Lambda\tilde\Lambda,\nonumber\\
&&(V_1,V_2)_{h,{\alpha}}=-\frac{1}{2}\bar D^2(V_1D_{\alpha}V_2-D_{\alpha}V_1V_2),\nonumber\\
&&(V,W)_h=D_{\alpha}VW^{\alpha}+\frac{1}{2}D^{\alpha}W_{\alpha}V,\nonumber\\
&&(\tilde W,W)_h=\tilde W^{\alpha}W_{\alpha},\nonumber\\
&&(\tilde V, V)=-\frac{1}{2}{\bar{D}}^2(\t VV),\nonumber\\
&&(W_1,W_2)_h=0,\quad (V, \tilde W)_h=0,
\end{eqnarray}
such that $\Lambda_1,\Lambda_2\in \Phi$; $V,V_1,V_2\in \Sigma$; $W,W_1,W_2\in\Theta$, 
$\tilde W\in \tilde \Theta$; $\tilde V\in \tilde \Sigma$; $\tilde\Lambda\in \tilde\Xi$. 

Now we start checking that this operation satisfies all necessary relations of 
homotopy associative algebra. The first relation is the Leibniz rule, i.e. 
\begin{eqnarray}
\tmd (a_1,a_2)_h=(\tmd a_1,a_2)_h+(-1)^{|a_1|}(a_1,\tmd a_2)_h.
\end{eqnarray}
Let us check it step by step:
\begin{eqnarray}
&&\tmd (\Lambda,V)_h=\frac{1}{2}\b D^2D_{\alpha}(\Lambda V)=\nonumber\\
&&\frac{1}{2}\b D^2(D_{\alpha}\Lambda V-\Lambda 
D_{\alpha}V)+\Lambda{\b D}^2 D_{\alpha}V=\nonumber\\
&&(\tmd\Lambda,V)_h+(\Lambda,\tmd V)_h,
\end{eqnarray}

\begin{eqnarray}
&&\tmd (\Lambda,W)_h=D^{\alpha}(\Lambda W_{\alpha})+\Lambda W_{\alpha}=\nonumber\\
&& D_{\alpha}\Lambda W^{\alpha}+\frac{1}{2}\Lambda D^{\alpha}W_{\alpha}+\Lambda W_{\alpha}+
\frac{1}{2}\Lambda D^{\alpha}W_{\alpha}=\nonumber\\
&&(\tmd\Lambda,W)_h+(\Lambda,\tmd W)_h,
\end{eqnarray}

\begin{eqnarray}
\tmd (V,W)_h=\b D^2(D_{\alpha}VW^{\alpha}+\frac{1}{2}VD^{\alpha}W_{\alpha})=\nonumber\\
\b D^2 D^{\alpha}VW_{\alpha}+\frac{1}{2}\b D^2VD^{\alpha}W_{\alpha}=\nonumber\\
(\tmd V,W)_h-(V,\tmd W)_h,
\end{eqnarray}

\begin{eqnarray}
\tmd (\Lambda,\tilde V)_h=\frac{1}{2}\b D^2(\Lambda \tilde V)=-\frac{1}{2}\Lambda \b D^2\tilde V+
\Lambda \b D^2\tilde V=
(\tmd\Lambda,\tilde V)_h+(\Lambda,\tmd\tilde V)_h.
\end{eqnarray}

Our next task is to verify that $(\cdot,\cdot)_h$ satisfies homotopy associativity relation:
\begin{eqnarray}
&& (a_1,(a_2,a_3)_h)_h-((a_1,a_2)_h,a_3)_h+\tmd (a_1,a_2,a_3)_h+ (\tmd a_1,a_2,a_3)_h+\nonumber\\
&&(-1)^{|a_1|}(a_1,\tmd a_2,a_3)_h+(-1)^{|a_1|+|a_2|}(a_1,a_2,\tmd a_3)_h=0,
\end{eqnarray}

where $(\cdot,\cdot,\cdot)_h$ is the trilinear operation we will derive below.

Let us proceed as above, checking associativity step by step:

\begin{eqnarray}
&&(\Lambda,(V_1,V_2)_h)_h-((\Lambda,V_1)_h,V_2)_h=\nonumber\\
&&-\frac{1}{2}\b D^2(\Lambda V_1D_{\alpha}V_2-\Lambda D_{\alpha}V_1 V_2)+
\frac{1}{4}\b D^2(\Lambda V_1D_{\alpha}V_2-V_1D_{\alpha}(\Lambda V_2))=\nonumber\\
&&-\frac{1}{4}\b D^2(\Lambda(V_1D_{\alpha}V_2-
D_{\alpha}V_1 V_2)+D_{\alpha}\Lambda V_1V_2)=\nonumber\\
&&-\tmd (\Lambda,V_1,V_2)_h-(\tmd\Lambda,V_1,V_2)_h-(\Lambda,\tmd V_1,V_2)_h+(\Lambda,V_1,\tmd V_2)_h,
\end{eqnarray}

\begin{eqnarray}
&&(V_1,(\Lambda,V_2)_h)_h-((V_1,\Lambda)_h,V_2)_h=\nonumber\\
&&-\frac{1}{4}\b D^2 (V_1D_{\alpha}(\Lambda V_2)-D_{\alpha}V\Lambda V_2)+\frac{1}{4}\b D^2 (V_1\Lambda D_{\alpha}V_2-D_{\alpha}(V_1\Lambda V_2)=\nonumber\\
&&-\frac{1}{2}\b D^2(V_1D_{\alpha}\Lambda V_2)=\nonumber\\
&&-\tmd (V_1,\Lambda,V_2)_h-(\tmd V_1,\Lambda,V_2)_h+(V_1,\tmd\Lambda,V_2)_h+(V_1,\Lambda,\tmd V_2)_h,
\end{eqnarray}

\begin{eqnarray}
&&(\Lambda_1,(\Lambda_2,V)_h)_h-((\Lambda_1,\Lambda_2)_h,V)_h=\frac{1}{4}(\Lambda_1\Lambda_2 V)-\frac{1}{2}(\Lambda_1\Lambda_2V)=\nonumber\\
&&-\frac{1}{4}\Lambda_1\Lambda_2 V=\nonumber\\
&&-\tmd (\Lambda_1,\Lambda_2,V)_h-(\tmd\Lambda_1,\Lambda_2,V)_h-(\Lambda_1,\tmd\Lambda_2,V)_h-(\Lambda_1,\Lambda_2,\tmd V)_h.
\end{eqnarray}
Since we have 
\begin{eqnarray}
((\Lambda_1,V)_h,\Lambda_2)_h=(\Lambda_1,(V,\Lambda_2)_h)_h,
\end{eqnarray}
we obtain 
\begin{eqnarray}
&&(V_1,V_2,V_3)_h=\frac{1}{6}\b D^2(V_1V_2D_{\alpha}V_3-2V_1(D_{\alpha}V_2)V_3+(D_{\alpha}V_1)V_2V_3)\nonumber\\
&&(\Lambda,V_1,V_2)_h=(V_1,V_2,\Lambda)_h=\frac{1}{12}\Lambda V_1 V_2\nonumber\\
&&(V_1,\Lambda,V_2)_h=\frac{1}{6}\Lambda V_1V_2
\end{eqnarray}

One can check that these expressions fit the formulas above.
Then we need to study the associativity relation involving 
$W$-terms, i.e.
\begin{eqnarray}
&&(\Lambda, (V,W)_h)_h-((\Lambda,V)_h,W)_h=\nonumber\\
&&\frac{1}{2}\Lambda D_{\alpha}VW^{\alpha}+\frac{1}{4}D^{\alpha}W_{\alpha}V\Lambda-\frac{1}{2}
D_{\alpha}(\Lambda V)W^{\alpha}-\frac{1}{4}\Lambda VD^{\alpha}W_{\alpha}=\nonumber\\
&&-\frac{1}{2}D_{\alpha}\Lambda VW^{\alpha}=\nonumber\\
&&-\tmd(\Lambda,V,W)_h-(\tmd\Lambda,V,W)_h-(\Lambda,\tmd V,W)_h+(\Lambda,V,\tmd W)_h,
\end{eqnarray}
where $(V_1,V_2, W)_h= \frac{1}{2}D_{\alpha}V_1V_2W^{\alpha}+\frac{1}{6}V_1V_2D_{\alpha}W^{\alpha}$ and $
(\Lambda, V, \t V)=\frac{1}{6}V_1V_2\t V$. Another terms involving $W$ are
\begin{eqnarray}
&&(V_1,(V_2,W)_h)_h-((V_1,V_2)_h,W)_h=\nonumber\\
&&-\frac{1}{2}\b D^2(\frac{1}{2}V_1(D_{\alpha}V_2W^{\alpha}+\frac{1}{2}D^{\alpha}W_{\alpha}V_2)+\frac{1}{2}\b D^2(V_1D_{\alpha}V_2-D_{\alpha}V_1V_2)W^{\alpha})=\nonumber\\
&&-\b D^2(\frac{1}{2}D_{\alpha}V_1V_2W^{\alpha}+\frac{1}{4}V_1D^{\alpha}W_{\alpha}V_2)\nonumber\\
&&-\tmd (V_1,V_2,W)_h-(\tmd V_1,V_2,W)_h+(V_1,\tmd V_2,W)_h-(V_1,V_2,\tmd W)_h,
\end{eqnarray}
 \begin{eqnarray}
&&(V_1, (W,V_2)_h)_h-((V_1, W)_h,V_2)_h)=\frac{1}{2}{\b D^2}(V_1(D^{\alpha}V_2W_{\alpha}+\frac{1}{2}D^{\alpha}W_{\alpha}V_2)+\nonumber\\
&&V_2(D^{\alpha}V_1W_{\alpha}+\frac{1}{2}D^{\alpha}W_{\alpha}V_1)=\nonumber\\
&&-\tmd (V_1,W,V_2)_h-(\tmd V_1,W,V_2)_h+(V_1,\tmd W,V_2)_h-(V_1,W,\tmd V_2)_h.
\end{eqnarray}
Here 
\begin{eqnarray}
&& (V_1,V_2,W)_h=(W,V_1,V_2)_h
=\frac{1}{2}D_{\alpha}V_1V_2W^{\alpha}+\frac{1}{6}V_1V_2D_{\alpha}W^{\alpha},\nonumber\\
&& (V_1,W,V_2)_h=-\frac{1}{2}(V_1D_{\alpha}V_2+D_{\alpha}V_1V_2)W^{\alpha}-\frac{1}{3}V_1D^{\alpha}W_{\alpha}V_2,\nonumber\\
&& (V_1,V_2,\t V)_h=(\t V, V_1,V_2)=\frac{1}{12}\b D^2(V_1V_2\t V),\nonumber\\
&& (V_1,\t V, V_2)=\frac{1}{6}\b D^2 (V_1\t V V_2).
\end{eqnarray}

%We can expand the equations \rf{em} and the symmetries up to the third order in V, W:
%\begin{eqnarray}
%&&D^{\alpha}W_{\alpha}+[D_{\alpha}V, W^{\alpha}]+\frac{1}{2}[[D_{\alpha}V,V],W^{\alpha}]+...=0\nonumber\\
%\nonumber
%&&W_{\alpha}+\b D^2 D_{\alpha}V-\frac{1}{3!}\b D^2[V,[V, D_{\alpha}V]]+...=0\nonumber\\
%&&V\to V+\Lambda-\b \Lambda +\frac{1}{2}[V,\Lambda+\b\Lambda]+
%\frac{1}{12}[V,[V,\Lambda-\b \Lambda]]+...,\nonumber\\
%&&W_{\alpha}\to W_{\alpha}+[W_{\alpha},\Lambda]
%\end{eqnarray}
Therefore from the calculations above one can obtain that algebraically up to the third order these equations look as follows:
\begin{eqnarray}
&&\tmd U+(U,U)_h+(U,U,U)_h+...=0,\nonumber\\
&&U\to U+\tmd\Lambda+(U,\Lambda)_h+\nonumber\\
&&(U,U,\Lambda)_h-(U,\Lambda,U)_h+(\Lambda,U,U)_h+...,
\end{eqnarray}
where $U$ is the element of the grading 1 from the complex \rf{fo} corresponding to the pair $(V,W)$.\\

%\section*{Appendix C}
%{\bf C1. Notations for $N=1$ 3D superspace}. 
\noindent{\bf B.3. Notations for N=2 3D superspace.} We work with three-dimensional space with coordinates $x^{\mu}$ and euclidean metric $\eta_{\mu\nu}=\rm{diag}(+1,+1,+1)$. The Dirac matrices are 
$(\gamma^{\mu})^{\beta}_{\alpha}=i(\sigma_1,\sigma_2,\sigma_3)$. We can raise and lower the corresponding spinor indices via $C^{\alpha\beta}$, i.e. $\xi^{\alpha}=C^{\alpha\beta}\xi_{\beta}$ and $\xi_{\alpha}=C_{\alpha\beta}\xi^{\beta}$, such that $C^{12}=-C_{12}=i$. 

The $N=2$ 3D superspace has two 2-component anticommuting coordinates $\theta^{\alpha}_1$ and $\theta^{\alpha}_2$. Therefore (this is similar to $N=1$ superspace) one can define complex coordinates $\theta^{\alpha}=\theta^{\alpha}_1-i\theta^{\alpha}_2$ and   $\b\theta^{\alpha}=\theta^{\alpha}_1+i\theta^{\alpha}_2$.  This allows to  define  superderivatives (cf. $N=1$ D=4 case):
\begin{eqnarray}
D_{\alpha}=\p_{\alpha}+\frac{i}{2}\bar{\theta}^{\beta}\p_{\alpha\beta}, \quad \b D_{\alpha}=\b \p_{\alpha}+\frac{i}{2}{\theta}^{\beta}\p_{\alpha\beta}
\end{eqnarray}
where $\p_{\alpha\beta}=\gamma^{\mu}_{\alpha\beta}\p_{\mu}$. One defines the (anti)chiral scalar fields via the 
familiar equation: $\b D_{\alpha}\Lambda(x,\theta)=0$, $D_{\alpha}\b \Lambda(x,\theta)=0$.    
For the calculations we will need the following commutation relations between superderivatives:
\begin{eqnarray}
\{D_{\alpha},\b D_{\beta}\}=i\gamma^{\mu}_{\alpha\beta}\p_{\mu}=i\p_{\alpha\beta}, \quad 
\{D_{\alpha},D_{\beta}\}=0, \quad \{\b D_{\alpha},\b D_{\beta}\}=0.
\end{eqnarray}
As well as in $N=1$ D=4 case, it is useful to introduce operators $D^2=C^{\alpha\beta}D_{\alpha}D_{\beta}$, $\b D^2=C^{\alpha\beta}\b D_{\alpha}\b D_{\beta}$.\\

%\noindent {\bf C2. Explicit calculations for the $A_{\infty}$-algebra of $N=2$ SUSY Chern-Simons theory.} 
%Here we verify explicitly some relations from the $A_{\infty}$-algebra related to the complex 
%$(SdR^{\cdot}, d_S)$. 

\end{document}